%% file: MLCask_main.tex
\documentclass[conference]{IEEEtran}
\IEEEoverridecommandlockouts

\input{terminology.tex}

\usepackage{tikz}

\usepackage{amsthm}
\theoremstyle{definition}
\newtheorem{defn}{Definition} 
\theoremstyle{plain}

\usepackage{xcolor}
\usepackage[ruled, linesnumbered]{algorithm2e}

\SetKwProg{Fn}{Function}{}{end}\SetKwFunction{PrUsCompat}{PruneUsingCompatibility}%
\SetKwProg{Fn}{Function}{}{end}\SetKwFunction{PrUsReusable}{PruneUsingReusable}%
\SetKwProg{Fn}{Function}{}{end}\SetKwFunction{ExecuteTree}{ExecuteTree}%
\SetKwProg{Fn}{Function}{}{end}\SetKwFunction{InitCompat}{InitCompatibilityTable}%
\SetKwRepeat{Do}{do}{while}
\SetAlgoLongEnd
\usepackage{multicol}

\newcommand{\icde}[1]{{\color{black}#1}} 
\newcommand{\shepherd}[1]{{\color{black}#1}} 

\usepackage{cite}
\usepackage{amsmath,amssymb,amsfonts}
\usepackage{graphicx}
\usepackage{subfigure}
\usepackage{textcomp}
\usepackage{comment}
\usepackage{multirow}
\usepackage{makecell}
\def\BibTeX{{\rm B\kern-.05em{\sc i\kern-.025em b}\kern-.08em
    T\kern-.1667em\lower.7ex\hbox{E}\kern-.125emX}}
\begin{document}

\title{\system{}: Efficient Management of Component Evolution in Collaborative Data Analytics Pipelines}

\author{
\IEEEauthorblockN{%
    \parbox{\linewidth}{\centering
%
{Zhaojing Luo\IEEEauthorrefmark{2}, 
Sai Ho Yeung\IEEEauthorrefmark{2}, 
Meihui Zhang\IEEEauthorrefmark{3}\textsuperscript{*}\thanks{* contact author}, 
Kaiping Zheng\IEEEauthorrefmark{2}, 
Lei Zhu\IEEEauthorrefmark{2},
Gang Chen\IEEEauthorrefmark{4}, 
Feiyi Fan\IEEEauthorrefmark{5}, 
Qian Lin\IEEEauthorrefmark{2}, 
Kee Yuan Ngiam\IEEEauthorrefmark{6},
Beng Chin Ooi\IEEEauthorrefmark{2}}%
}
}
\\
\IEEEauthorrefmark{2} National University of Singapore \hspace{2mm}
\IEEEauthorrefmark{3} Beijing Institute of Technology \hspace{2mm}
\IEEEauthorrefmark{4} Zhejiang University \\
\IEEEauthorrefmark{5} ICTCAS \hspace{2mm}
\IEEEauthorrefmark{6} National University Health System, Singapore \\
\{zhaojing,yeungsh,kaiping,zhu-lei,ooibc\}@comp.nus.edu.sg \hspace{2mm}
meihui\_zhang@bit.edu.cn \hspace{2mm} 
cg@zju.edu.cn \\
fanfeiyi@ict.ac.cn  \hspace{2mm}
linqian.qian@hotmail.com \hspace{2mm} 
kee\_yuan\_ngiam@nuhs.edu.sg \\
}

\maketitle

\begin{abstract}
With the ever-increasing adoption of machine learning for data analytics, maintaining a machine learning pipeline is becoming more complex as both the datasets and trained models evolve with time. 
In a collaborative environment, the changes and updates due to pipeline evolution often cause cumbersome coordination and maintenance work, raising the costs and making it hard to use.
Existing solutions, unfortunately, do not address the version evolution problem, especially in a collaborative environment where non-linear version control semantics are necessary to isolate operations made by different user roles. 
The lack of version control semantics also incurs unnecessary storage consumption and lowers efficiency due to data duplication and repeated data pre-processing, which are avoidable.

In this paper, we identify two main challenges that arise during the deployment of machine learning pipelines, and address them with the design of versioning for an end-to-end analytics system \system{}.
The system supports multiple user roles with the ability to perform \git{}-like branching and merging operations in the context of the machine learning pipelines. 
We define and accelerate the metric-driven merge operation by pruning the pipeline search tree using reusable history records and pipeline compatibility information. 
Further, we design and implement the prioritized pipeline search, which gives preference to the pipelines that probably yield better performance.
The effectiveness of \system{} is evaluated through an extensive study over several real-world deployment cases.
The performance evaluation shows that the proposed merge operation is up to 7.8x faster and saves up to 11.9x storage space than the baseline method that does not utilize history records.
\end{abstract}


\section{Introduction}
In many real-world machine learning (ML) applications, new data is continuously fed to the ML pipeline.
Consequently, iterative updates and retraining of the analytics components become essential, especially for applications that exhibit significant concept drift behavior where the trained model becomes inaccurate as time passes.
Consider healthcare applications~\cite{dai2018fine,zheng2020tracer} as an example in which hospital data is fed to data analytics pipelines~\cite{guzzetta2010machine,luo2018adaptive} on a daily basis for various medical diagnosis predictions.
The extracted data schema, pre-processing steps, analytics models 
are highly volatile~\cite{lee2017big,zheng2017resolving} due to the evolution of the dataset, 
leading to a series of challenges.
First, to ensure quality satisfaction of the analytics models, the pipeline needs to be retrained frequently to adapt to the changes, which costs a lot of storage and time~\cite{wang2016database,cai2019model,van2017versioning}.
Second, the lengthy pipeline and computer cluster environment cause the asynchronous pipeline update problem, because different components may be developed and maintained by different users. 
Third, the demand for retrospective research on models and data from different time periods further complicates the management of massive pipeline versions.

To address the aforementioned challenges, version control semantics~\cite{huang2017rpheus, maddox2016decibel, van2017versioning, wang2018forkbase} need to be introduced to the ML pipeline. 
Current pipeline management systems either do not explicitly consider the version evolution, or handle versioning by merely archiving different versions into distinctive disk folders so that different versions will not conflict with or overwrite each other. 
The latter approach not only incurs huge storage and computation overhead, but also fails to describe the logical relationship between different versions.
In this paper, we first elaborate on the common challenges in data analytics applications and formulate version control semantics in the context of ML pipeline management.
We then present a design of \git{}-like end-to-end ML life-cycle management system, called \system{}, and its version control support. 
\system{} facilitates collaborative component updates in ML pipelines, where components refer to the computational units in the pipeline such as data ingestion methods, pre-processing methods, and models. 
The key idea of \system{} is to keep track of the evolution of pipeline components together with the inputs, execution context, outputs, and the corresponding performance statistics. 
By introducing the non-linear version control semantics~\cite{huang2017rpheus, maddox2016decibel, wang2018forkbase} to the context of ML pipelines, \system{} can achieve full historical information traceability with the support of branching and merging. 
Further, we propose two methods in \system{} to prune the pipeline search tree and reuse materialized intermediate results to reduce the time needed for the metric-driven merge operation. 
Lastly, 
to minimize the cost of the merge operation for divergent ML pipeline versions, we devise multiple strategies in \system{} that prioritize the search for the more promising pipelines ranked based on the historical statistics. 

The main contributions of this paper can be summarized as follows:
\begin{itemize}
\item We identify two key challenges of managing asynchronous activities between agile development of analytics components and retrospective analysis. 
Understanding these challenges provides the insights for efficiently managing the versioning of ML pipelines. 
\item We present the design of an efficient system \system{}, with the support of non-linear version control semantics in the context of ML pipelines.
\system{} can ride upon most of the mainstream ML platforms to manage component evolution in collaborative ML pipelines via branching and merging. 
\item We propose two search tree pruning methods in \system{} to reduce the candidate pipeline search space in order to improve system efficiency under the non-linear version control semantics.
We further provide a prioritized pipeline search strategy in \system{} that looks for promising but suboptimal pipelines with a given time constraint.
\item \icde{We have fully implemented \system{} for deployment in a local hospital.
Experimental results on diverse real-world ML pipelines demonstrate \system{} achieves better performance than baseline systems, ModelDB~\cite{vartak2016modeldb} and MLflow~\cite{zaharia2018accelerating}, in terms of storage efficiency and computation reduction.} 
 
\end{itemize}

The remainder of the paper is structured as follows. 
Section~\ref{sec: challenge} introduces the background and motivation of introducing version control semantics to machine learning pipelines.
Section~\ref{sec: system design} presents the system architecture of \system{}.
Section~\ref{sec: version control semantics} presents the version control scheme of \system{} and Section~\ref{sec: supporting non-linear version history} introduces the support of non-linear version history in \system{}.
The optimization of merge operations is presented in Section~\ref{sec: optimizing version operations}. 
Experimental results and discussions on the prioritized pipeline search are presented in Section~\ref{sec: evaluation}. 
\icde{We share our experience on the system deployment in Section~\ref{sec: discussion}.}
Related work is reviewed in Section~\ref{sec: related work} and we conclude the paper in Section~\ref{sec: conclusions}.


\section{Challenges of Supporting Data Analytics Applications}
\label{sec: challenge}

In many real-world data analytics applications, not only data volume keeps increasing, but also analytics components undergo frequent updates. 
A platform that supports intricate activities of data analytics has to address the following two key challenges.

\textbf{(C1) Frequent retraining.}
Many real-world data analytics applications require frequent retraining since concept drift is a common phenomenon. 
For instance, in the computer cluster of \hospital{}\footnote{National University Health System (\hospital{}) consists of four public hospitals, seven polyclinics, and a number of research institutes and academic units.} hospital, there are around $800$ to $1200$ inpatients at any given time and the number of newly admitted patients each day is around $150$. 
Given this dynamic environment, retraining models by using new patient data from time to time is essential for delivering accurate predictions. 
Currently, the existing workflow needs to rerun every component for each retraining, which is time consuming and resource intensive. 
Meanwhile, different pipeline versions are archived into separate folders, which leads to huge storage consumption.
To overcome the aforementioned resource problems, a mechanism is needed to identify the component that does not need to be rerun for efficient pipeline management.
Furthermore, a component's output could be just partially different from the output of its previous version; hence, archiving them into separate folders does not resolve the storage redundancy.

\textbf{(C2) Asynchronous pipeline component update and merge.}
\icde{As expected for collaborative analytics, concurrent updates of a pipeline introduce both consistency and maintenance issues.
First, the asynchronous component update by different users may cause the potential failure of the entire pipeline when two incompatible updated components are combined.
Second, we should consider the fundamental difference between software engineering and building ML pipelines: ML pipeline development is metric-driven, rather than feature-driven. 
For building ML pipelines, data scientists typically pursue pipeline performance, and different branches are used for iterative trials. 
They often create different branches for iterative trials to improve individual components of the pipeline.
In contrast, software engineers merge two branches because the features developed on the merging branches are needed.
}

\icde{In the context of ML pipeline, simply merging two branches with the latest components does not necessarily produce a pipeline with improved performance, because the performance of the whole pipeline depends on the interaction of different components.
Therefore, developing ML pipelines through the collaboration of independent teams that consist of dataset providers (data owners), model developers, and pipeline users is challenging but necessary for better exploitation of individual knowledge and effort.
Consequently,
we have to address the issue of merging the pipeline updates from different user roles and searching for the best component combination among a massive amount of possible combinations of updates based on performance metrics.
}

In order to address the aforementioned challenges, version control semantics are incorporated into our end-to-end system \system{} as follows. 
By leveraging the version history of pipeline components and workspace, skipping unchanged pre-processing steps is realized in Section~\ref{sec: version control semantics} to address (C1), and non-linear version control semantics and merge operation are realized in Sections~\ref{sec: supporting non-linear version history} and \ref{sec: optimizing version operations} to address (C2).

\section{System Architecture of \system{}}
\label{sec: system design}
\begin{figure}[!ht]
  \centering
  \includegraphics[width=.99\linewidth]{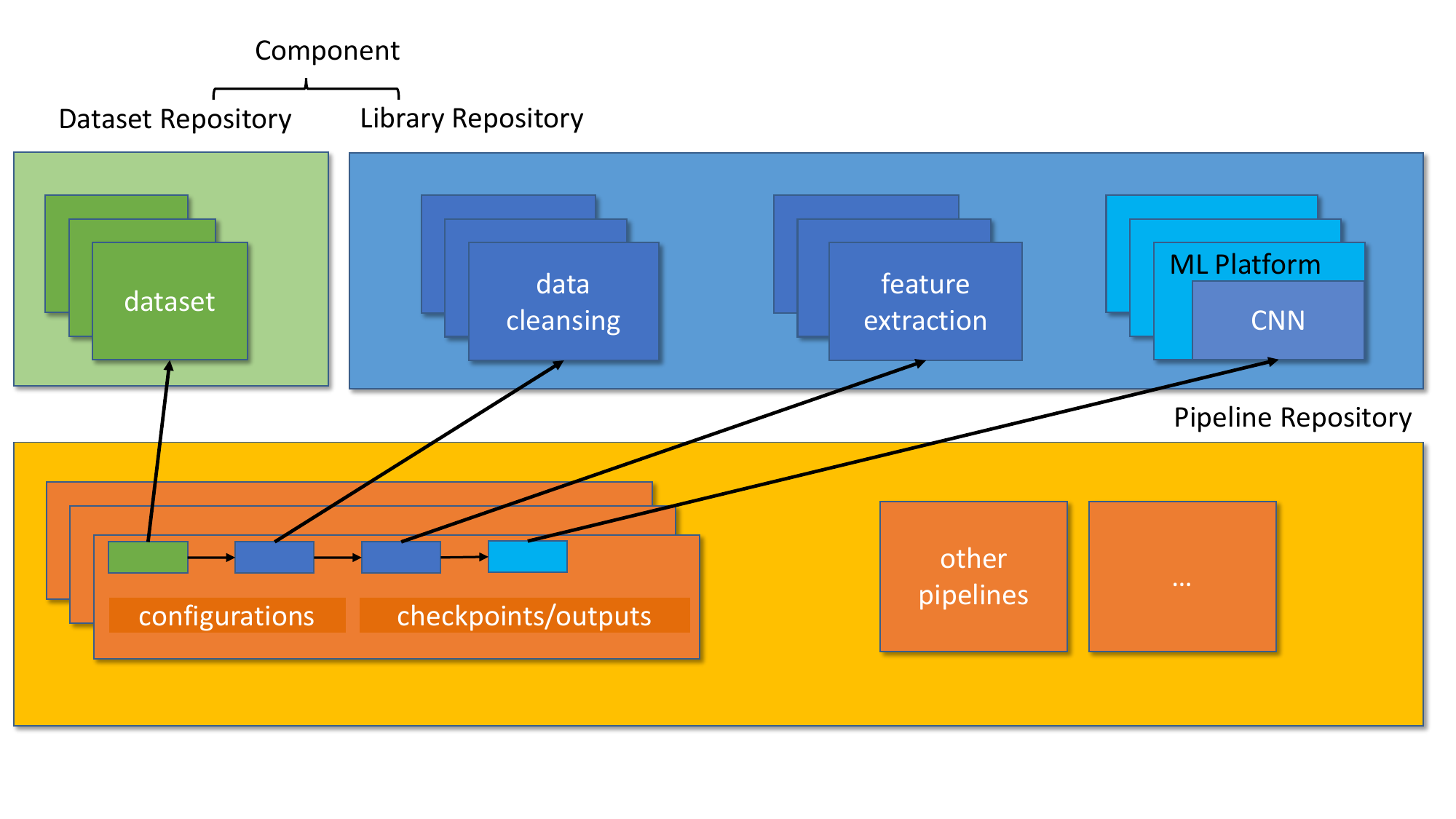}
  \caption{\icde{The architecture of \system{} for supporting collaborative pipeline development with version control semantics.}}
  \label{fig:gemini_arch}
\end{figure}

In this section, we introduce the system architecture of the ML life-cycle management system \system{}, which facilitates collaborative development and maintenance of ML pipelines. 
\system{} 
provides version control, stores evaluation results as well as provenance information, and records the dependency of different components of the pipelines.
The architecture of \system{} is illustrated in Fig.~\ref{fig:gemini_arch}.

In general, we abstract an ML life-cycle with two key concepts: component and pipeline.


\textbf{Component:} 
A component refers to any computational units in the ML pipeline, including \textit{datasets}, pre-processing methods, and ML models. 
In particular, we refer \textit{library} to either a pre-processing method or an ML model.

A dataset is an encapsulation of data which could either be a set of data files residing in a local/server side, or defined by the combination of database connection configurations and the associated data retrieval queries.
A dataset contains a mandatory metafile that describes the encapsulation of data and a series of optional data files.

\icde{
A library consists of a mandatory metafile and several executables.
It performs data pre-processing tasks or deep analytics.
The mandatory metafile describes the entry point, inputs and outputs, as well as all the essential hyperparameters in running the library. 
For a library of ML model training, the commonly used hyperparameters could be the learning rate and the maximum number of iterations.
}
In our implementation, we employ Apache SINGA~\cite{ooi2015singa}, a distributed deep learning system as the backend for training deep learning models.
Besides Apache SINGA, \system{} also readily works with other systems such as TensorFlow\footnote{https://www.tensorflow.org} or PyTorch\footnote{https://pytorch.org/} as long as the interface is compatible with the ML pipeline. 


\textbf{Pipeline:} 
A pipeline is the minimal unit that represents a ML task. 
When a pipeline is created with the associated components, the references to the components are recorded in the pipeline metafile. 
\icde{
A pipeline metafile describes the entry point of the pipeline and the order of the pipeline components such as data cleansing and the ML model.
Since the input/output schemas of the components are subject to change during the commits in the development process, the metafile of the components should be separated from the metafile of the pipeline. 
}
Once a pipeline is fully processed, all its component outputs are archived for future reuse, with their references logged into the pipeline metafile. 
Considering that a single dataset or library may be used by multiple pipelines, we design a \textit{dataset repository} and a \textit{library repository} to store different versions of datasets and libraries respectively, which are shared by all the pipelines in order to reduce storage costs.
A \textit{pipeline repository} is also introduced to record the version updates of all the pipelines. 

\textbf{Running Example:} 
To appreciate the discussion in the rest of the paper, without loss of generality, we exemplify an ML pipeline, as shown in Fig.~\ref{fig:gemini_arch}, which consists of datasets, data cleansing, feature extraction, and a convolutional neural network (CNN) model. 
This ML pipeline is used to predict whether a patient will be readmitted into the hospital within 30 days after discharge.

\section{Version Control Semantics}
\label{sec: version control semantics}

\subsection{Preliminaries}
\label{sec: preliminaries}
We use Directed Acyclic Graph (DAG) to formulate an ML pipeline as follows: 

\begin{defn}[\textbf{ML Pipeline}]
An ML pipeline $p$ with components $f_{i} \in \mathcal{F}$ is defined by a DAG $G = (\mathcal{F}, \mathcal{E})$, where each vertex represents a distinct component of $p$ and each edge in $\mathcal{E}$ depicts the successive relationship (i.e., direction of data flow) between its connecting components. 
\end{defn}

\begin{defn}[\textbf{Pipeline Data Flow}]
For a component $f \in \mathcal{F}$, let $suc(f)$ and $pre(f)$ be the set of succeeding and preceding components of $f$ respectively. 
Correspondingly, given components $f_{i}, f_{j} \in \mathcal{F}$ and a data flow $e_{ij} \in \mathcal{E}$ from $f_{i}$ to $f_{j}$, we have $f_{j} \in suc(f_{i})$ and $f_{i} \in pre(f_{j})$.
\end{defn}

\begin{defn}[\textbf{Pipeline Component}]
A pipeline component $f_i$ with the type of \emph{library} can be viewed as a transformation: $y = f_i(x | \theta_i)$, where $x$ is the input data of $f_i$, $\theta_i$ is the component's parameters, and $y$ denotes $f_i$'s output.
\end{defn}

\begin{defn}[\textbf{Component Compatibility}]
A pipeline component $f_{j}$ is compatible with its preceding component $f_{i} \in pre(f_{j})$ if $f_{j}$ can process the output by component $f_{i}$ correctly. 
\end{defn}

\subsection{Version Control for Pipeline Components}
\label{sec: version control of pipeline components}
A \emph{semantic version}\footnote{https://semver.org/} in \system{} is represented by an identifier: 
\verb|branch@schema.increment|, where $branch$ represents the \git{}-like branch semantics, $schema$ denotes the output data schema, and $increment$ represents the minor incremental changes that do not affect the output data schema. 

We use the notation: \verb|<feature_extract, master@0.1>| to denote a component named \verb|feature_extract| and its corresponding semantic version. 
The representation indicates that the component has received one incremental update and there is no output data schema update yet. 
For components on its \verb|master| branch, we simplify the representation to the following form: \verb|<feature_extract, 0.1>|. 
The initial version of a committed library is set to \verb|0.0|. 
Subsequent commits only affect the $increment$ domain if $schema$ is not changed. 
In this paper, we assume that the output data schema is the only factor that determines the compatibility between $f_i$ and $f_j$.
\icde{
Specifically, if the output data schema of $pre(f_i)$ changes, $f_i$ should perform at least one $increment$ update to ensure its compatibility with $pre(f_i)$.
}

For a library component, the update to $schema$ is explicitly indicated by the library developer in the library metafile\footnote{Library metafile is a configuration file written by the library developer, which records meta information of the library. It resides in the root folder of the library.}. 
For a dataset component, 
we propose that the data provider uses the schema hash function to map the $schema$ from data. For data in relational tables, all the column headers are extracted, standardized, sorted, and then concatenated into a single flat vector. 
Consequently, a unique $schema$ can be generated by applying a hash function such as SHA256 on the vector obtained.
Note that there are many methods available in the literature on the hash function optimization and this is not the focus of \system{}. 
\icde{For non-relational data, we can adopt the meta information which determines whether the dataset is compatible with its succeeding libraries being used, e.g., shape for image datasets, vocabulary size for text datasets, etc.}

Managing linear version history in ML pipeline has been well studied in literature~\cite{van2017versioning}. 
However, existing approaches cannot fulfill the gap when non-linear versioning arises, which is common in ML pipelines where multiple user roles are involved.
To tackle this problem, we develop the \system{} system to support non-linear version management in collaborative ML pipelines.

\section{Supporting Non-linear Version Control}
\label{sec: supporting non-linear version history}

\begin{figure}[!t]
  \centering
  \includegraphics[width=.9\linewidth]{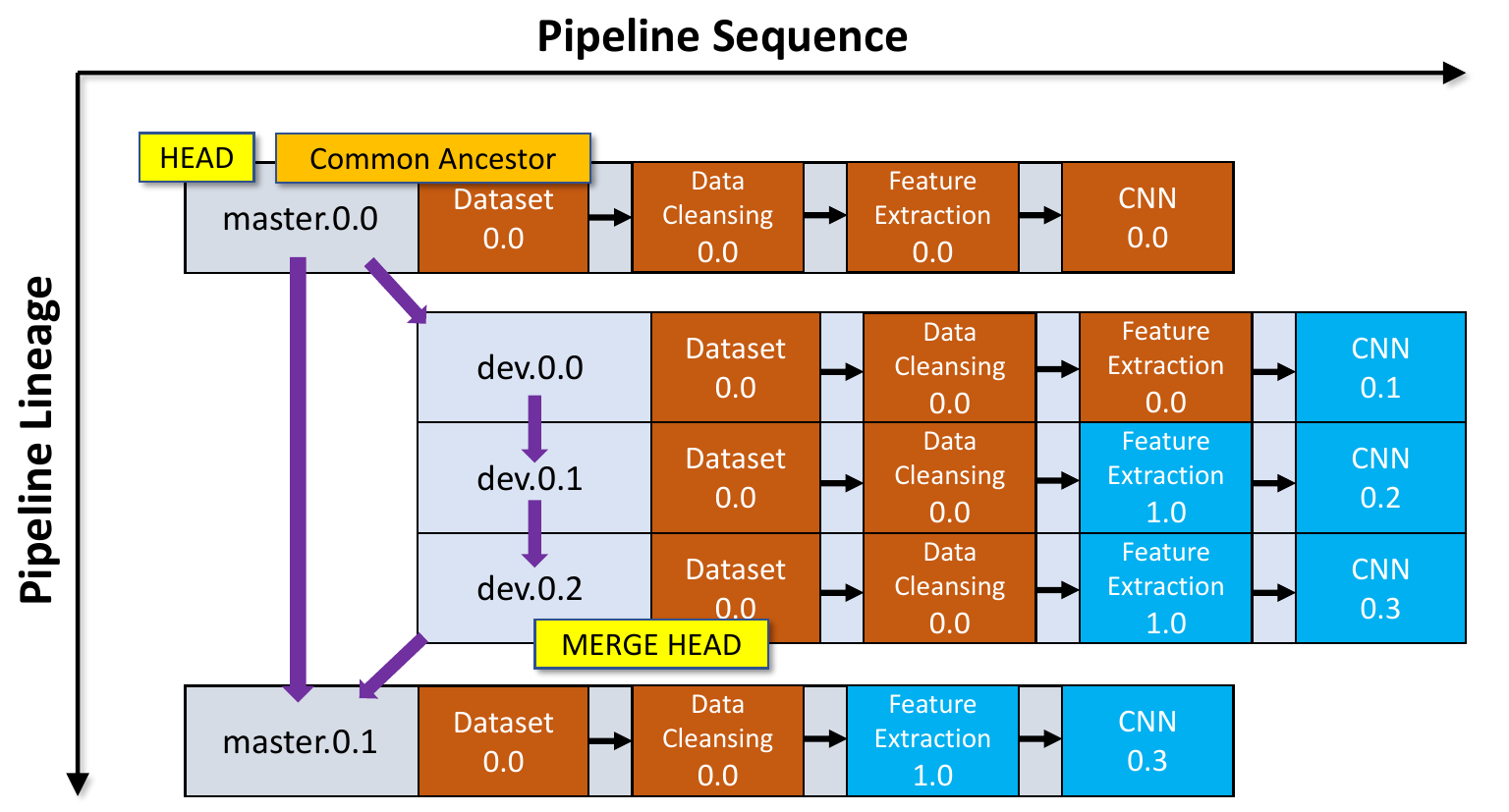}
  \caption{\system{} pipeline branching and merging without conflicts.}
  \label{fig:branch-ffmerge}
\end{figure}

\begin{figure*}[!t]
  \centering
  \includegraphics[width=0.7\linewidth]{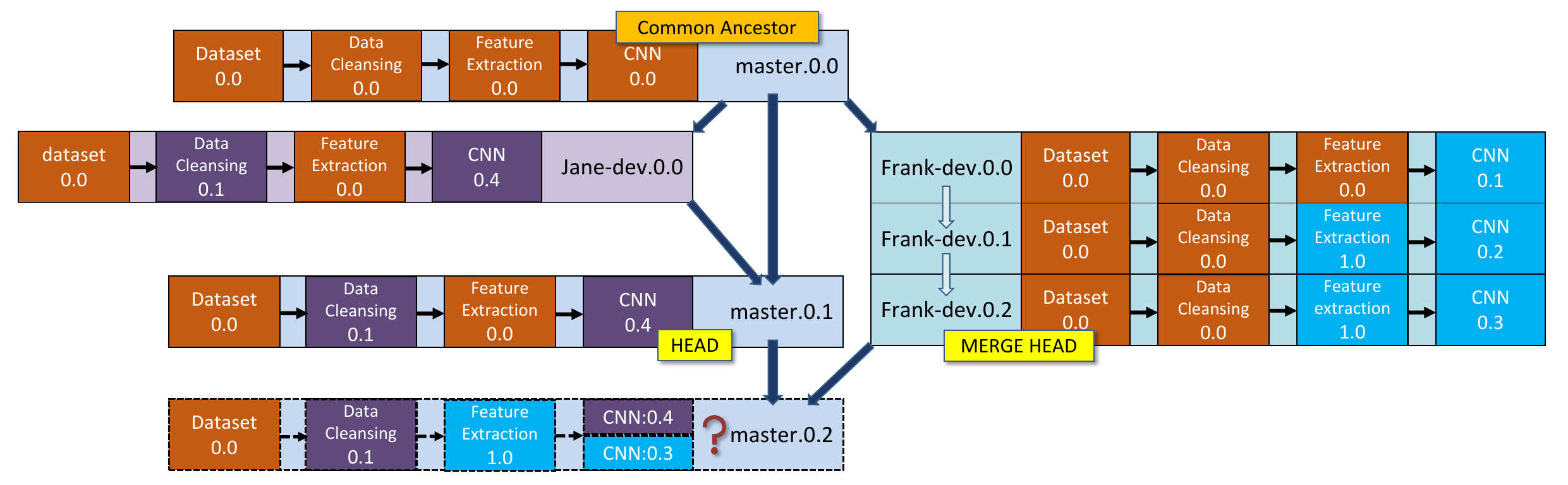}
  \caption{\system{} pipeline branching and merging with conflicts.}
  \label{fig:branch-merge}
\end{figure*}

We use the pipeline shown in Fig.~\ref{fig:branch-ffmerge} to illustrate how \system{} achieves \emph{branch} and \emph{merge} operations to support non-linear version history.
The example pipeline fetches data from a hospital dataset, followed by data cleansing and feature extraction, and eventually feeds the extracted data into a CNN model to predict how likely a specific patient will be readmitted in $30$ days.

\textbf{Branch}: In the collaborative environment, committing on the same branch brings in complications in the version history. 
It is thus desirable to isolate the updates made by different user roles or different purposes.
To address this issue, \system{} is designed to support \emph{branch} operations on every pipeline version. 
As shown in Fig.~\ref{fig:branch-ffmerge}, the \verb|master| branch remains unchanged before the \emph{merge} if all updates are committed to the \verb|dev| branch. 
By doing so, the isolation of a stable pipeline and development pipeline can be achieved.

\textbf{Merge}:
The essence of merging a branch to a base branch is to merge the commits (changes) that happened on the merging branch to the base branch. 
By convention, we term the base branch as \verb|HEAD| and the merging branch as \verb|MERGE_HEAD|. 

For the simplest case shown in Fig.~\ref{fig:branch-ffmerge}, the \verb|HEAD| does not contain any commits after the common ancestor of \verb|HEAD| and \verb|MERGE_HEAD|, which is constrained by the fast-forward merge. 
For the fast-forward merge, \system{} duplicates the latest version in \verb|MERGE_HEAD|, changes its branch to \verb|HEAD|, creates a new commit on \verb|HEAD|, and finally sets its parents to both \verb|MERGE_HEAD| and \verb|HEAD|.
%
However, if any commits happen on the \verb|HEAD| after the common ancestor, the resulting conflicts may become an issue. 
An example is illustrated in Fig.~\ref{fig:branch-merge}, in which the component CNN is changed on \verb|HEAD| before the merge. 


In terms of the merge operation in this scenario, a na\"ive strategy is to select the latest components to form the merging result.
\icde{
However, the na\"ive strategy is problematic for two reasons: (i) incompatibility, and (ii) sub-optimal pipeline. 
For the first reason, merging two different pipelines could lead to incompatibility issues between the components. For instance}, 
\verb|<CNN, 0.4>| in Fig.~\ref{fig:branch-merge} is not compatible with \verb|<feature_extract, 1.0>|
\icde{in their input/output schemas, which is reflected by the major version number of the feature extraction}. 

\icde{
For the second reason, the na\"ive strategy does not guarantee optimal performance due to complex coupling among pipeline components.
In the two branches HEAD and MERGE HEAD of Fig.~\ref{fig:branch-merge}, the three updated components Data Cleansing, Feature Extraction, and CNN are better than their old counterparts when they are evaluated separately.
However, the performance of the new pipeline that incorporates updates from both branches is unknown until it is actually evaluated.
For example, the version of Feature Extraction has been updated to 1.0 in the MERGE HEAD, but it is unknown that the updated CNN 0.4 in the HEAD can achieve good accuracy when it applies the new Feature Extraction 1.0. 
We should consider the performance of a pipeline in totality, instead of the individual performance of each component. 
The solution space is thus dependent on the pipeline search space which is typically huge and could have multiple local optima.
}

These observations motivate us to redefine the merge operation for the ML pipeline. 
Our assumption is that in \system{}, different users collaboratively update the pipeline in order to improve the performance, which is measured by a specific metric.
To be specific, we propose the \textit{metric-driven merge operation}, which aims to select an ML pipeline with optimal performance based on past commits made on \verb|HEAD| and \verb|MERGE_HEAD| referring to their common ancestor.

To this end, we first define the search space for selecting the optimal ML pipeline and then represent the search space using a \emph{pipeline search tree}.
\icde{
The search space involves all the available component versions developed starting from the common ancestors towards the HEAD and MERGE HEAD. 
Since the purpose of the development is to improve the pipeline at the common ancestor, the versions before the common ancestor are not considered since they could be outdated or irrelevant to the pipeline improvement. 
This leads to much reduction of computation time.
}

In Fig.~\ref{fig:branch-merge}, the component CNN has experienced $5$ versions of updates based on their common ancestor, and as a consequence, all these $5$ versions will be evaluated by the process of pipeline merge. 
Here we formalize the definition of ``all available component versions'' with respect to the concept of component search space.
Given $f_i$ is a component of pipeline $p$, the search space of $f_i$ on $p$'s branch $b$ is defined by:
\[
\mathcal{S}_b(f_i) = \{v(f_i|p) | p \in \mathcal{P}_b\}, 
\]
where $v(f_i | p)$ is the version of $f_i$ in pipeline $p$, $\mathcal{P}_b$ is the set of pipeline versions on the branch $b$. When merging two branches, \emph{component search space} of $f_i$ can be derived by:
\[
\mathcal{S}(f_i) = \mathcal{S}_{\verb|MERGE_HEAD|}(f_i) \cup \mathcal{S}_{\verb|HEAD|}(f_i).
\]

For data cleansing component in Fig.~\ref{fig:branch-merge}, its component search space contains two versions, namely:
\begin{verbatim}
  <data_cleanse, 0.0>, <data_cleanse, 0.1>
\end{verbatim}

To facilitate the search for the optimal combination of pipeline component updates, we propose to build a pipeline search tree using Algorithm~\ref{alg:generatetree} to represent all possible pipelines.
In Algorithm~\ref{alg:generatetree}, $S(f_i)$ denotes the component search space of $f_i$, $N_f$ is the number of pipeline components, and $tree$ is the returned pipeline search tree.

\begin{figure}[!t]
  \centering
  \includegraphics[width=.9\linewidth]{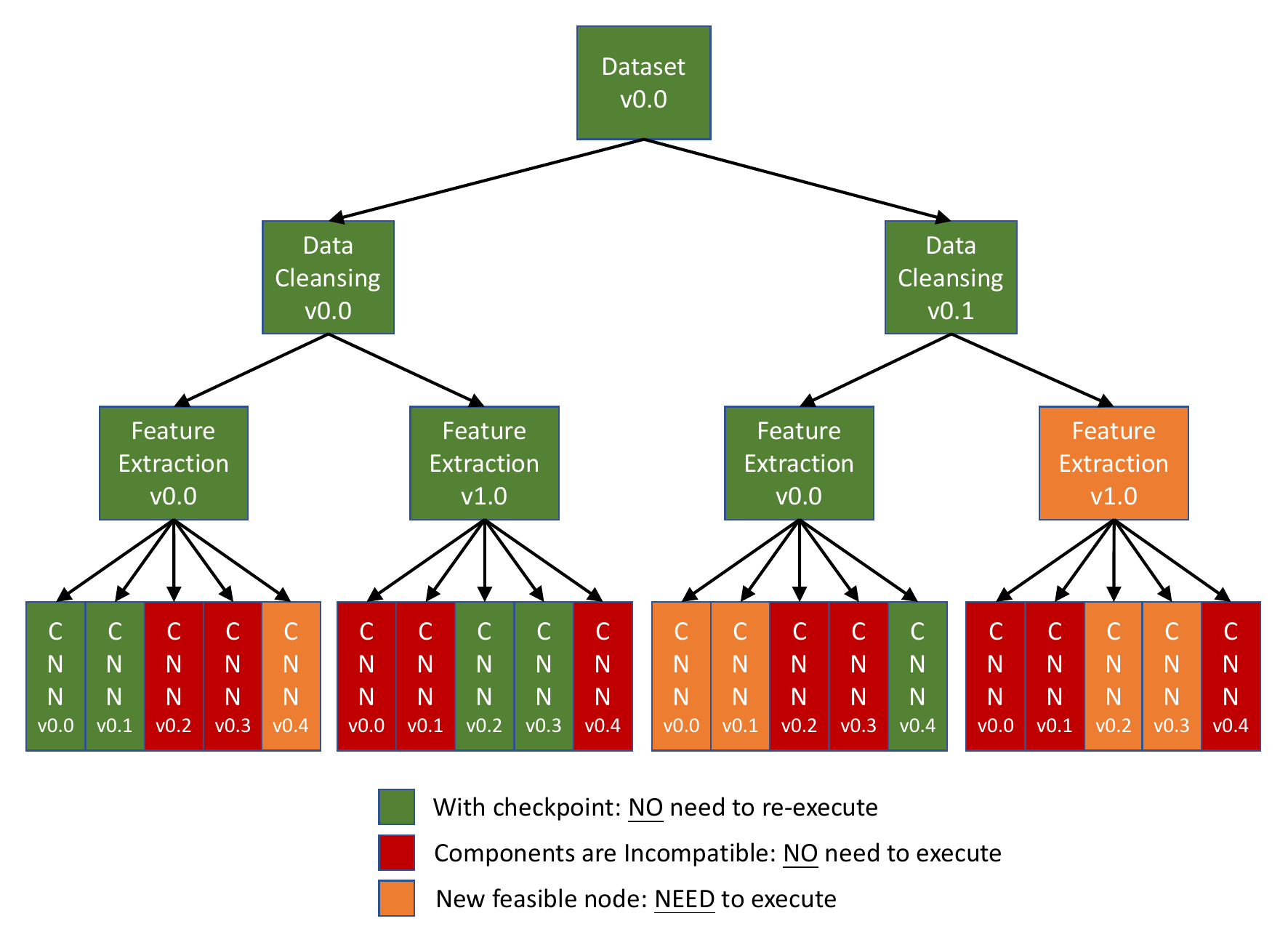}
  \caption{An example pipeline search tree built on version history.}
  
  \label{fig:pipeline-search-tree}
\end{figure}

\icde{
Fig.~\ref{fig:pipeline-search-tree} illustrates an example of a pipeline search tree, which is generated according to the merge operation in Fig.~\ref{fig:branch-merge} between the two branches HEAD and MERGE HEAD.
}
Every \verb|TreeNode| records the reference to a set of child nodes, its corresponding pipeline component, an execution status flag, and the reference to the component's output.
\icde{
There are three types of nodes denoted with different colors: The nodes in green color already have checkpoints in the development history starting from the common ancestor as depicted in Fig.~\ref{fig:branch-merge}. 
The nodes in red color are not executable due to the incompatibility between pipeline components, which are determined by the compatibility information introduced in Section~\ref{sec: prun} together with the semantic version rule in Section~\ref{sec: version control of pipeline components}. 
Finally, the nodes in orange, called feasible nodes, are the remaining nodes that need to be executed.
The nodes in red and green colors will be further elaborated in Sections~\ref{sec: prun} and \ref{sec:reuseable} respectively.
}

\begin{algorithm}[t]
\scriptsize
\caption{Pipeline search tree construction.}
\label{alg:generatetree}
\SetAlgoLined
\textbf{Input}: $S(f_i)$, $N_f$\\
\textbf{Output}: tree\\
 tree = TreeNode(component = virtual\_root, executed = True)\;
 \For{$i \gets 0$ \KwTo $N_f$}{
    fSet = $S(f_i)$\;
    parentNodes = tree.getNodeAtLevel(i)\;
    \ForEach{node $\in$ parentNodes}{
        \ForEach{f $\in$ fSet}{
            node.children.add(TreeNode(component = f, executed = False))
        }
    }
 }
\end{algorithm}

All possible pipelines can be obtained by enumerating all paths from the root to the leaves. The set of all the enumerated pipelines is termed as \emph{pre-merge pipeline candidates}, and is denoted as $\mathcal{P}_{candidate}$. The merged result can be defined by:
\[
p_{merged} = \mathop{\arg\max}_{p} \{ score(p) \mid p \in \mathcal{P}_{candidate}\},
\]
where $score(p)$ denotes the metric score that measures the performance of a pipeline. 
The form of the score function is dependent on the performance metrics used by the pipeline. For example, we can use $score = \frac{1}{MSE}$ as a score function for a pipeline whose performance metric is the mean squared error (MSE). If there are different metrics for evaluation, \system{} generates different optimal pipeline solutions for different metrics so that users could select the most suitable one based on their preference.

\section{Optimizing Merge Operations}
\label{sec: optimizing version operations}
In this section, we present optimizations to improve the efficiency of the \emph{merge} operations in \system{}.
\icde{The non-triviality of the merge operation lies in the huge search space for the optimal pipeline and how to exclude the incompatible pipelines.} 
For a pipeline with $N_f$ components, the upper bound of the number of the possible pipeline candidates is given by $
\prod_{i = 1}^{N_f} N(\mathcal{S}(f_i))$, where $N(\mathcal{S}(f_i))$ denotes the number of elements in set $\mathcal{S}(f_i)$. Therefore, the number of pipeline candidates increases dramatically when the number of past commits increases, which may render the \emph{merge} operation extremely time-consuming. 

Fortunately, among a large number of pipeline candidates, those with incompatible components can be safely excluded.
Further, if a component of the pipeline candidate was executed before, it does not need to be executed again since its output has already been saved and thus can be reused.
Motivated by these two observations, we propose two tree pruning methods to accelerate the merge operation in \system{}.

\subsection{Pruning Merge Tree using Component Compatibility Information}
\label{sec: prun}
When the $schema$ of a pipeline component changes, its succeeding components have to be updated accordingly. 
By leveraging the constraints on component compatibility, we can avoid enumerating the pipelines that are destined to fail in execution. 

\icde{We continue to use the version history as illustrated in Fig.~\ref{fig:branch-merge} and its corresponding pipeline search tree in Fig.~\ref{fig:pipeline-search-tree} to exemplify the idea and show the compatibility information.}
The succeeding components of feature extraction can be divided into two sets based on compatibility:
\begin{itemize}
    \item \{\verb|<CNN, 0.0>|, \verb|<CNN, 0.1>|, \verb|<CNN, 0.4>|\}\\ following \verb|<feature_extract, 0.0>|;
    \item \{\verb|<CNN, 0.2>|, \verb|<CNN, 0.3>|\}\\following \verb|<feature_extract, 1.0>|;
\end{itemize}

\icde{
In Fig.~\ref{fig:pipeline-search-tree}, the nodes in red are not compatible with their parent nodes. 
By pruning all those nodes, the size of the pre-merge pipeline candidate set can be reduced to half of its original size.
}

In practice, a compatibility look-up table (LUT) is evaluated based on the pipelines' version history to support the pruning procedure. 
Firstly, given a component, all its versions on the \verb|HEAD| and \verb|MERGE_HEAD| are enumerated. 
Secondly, for every version of the given component, we find its compatible succeeding component versions. 
Finally, we make the compatible component pairs in 2-tuple and fill the LUT with 2-tuple.   
Once the compatibility LUT is obtained, it can be used to prune the pipeline search tree. 
Pruning incompatible pipelines not only narrows the search space, but also solves the asynchronous pipeline update problem in non-linear version control semantics because all incompatible pipelines are pruned. 
This procedure can be integrated with depth-first-traversing the pipeline search tree which will be introduced in Section~\ref{sec:reuseable}.

\subsection{Pruning Merge Tree using Reusable Output}
\label{sec:reuseable}
Apart from pruning the pipeline search tree by inspecting the pipeline component compatibility, the reusable output could be utilized as a pruning heuristic to avoid unnecessarily repeated computation. 
The key to achieve this is to precisely identify the common procedures between pipeline versions so that the execution of the new pipeline can be based on the differences in components between pipeline versions rather than always starting from scratch. 

An important feature of a pipeline search tree is that every node has only one parent node, which means the nodes sharing the same parent node also share the same path to the tree root. 
Once a node is executed, all its children nodes will benefit from reusing its output. 
Therefore, pruning the pipeline search tree can be implemented in the following two steps.

First, we mark the node with an execution status using the previously trained pipelines in the commit history. 
As illustrated in Fig.~\ref{fig:pipeline-search-tree}, the nodes in green are examples for this case. 
Note that a reference to the component's output is recorded in the node object for future reuse. 

Second, we mark the node with an execution status when traversing and executing every node's corresponding component on the pipeline search tree. 
Depth-first traversal is suitable for the problem because it guarantees that once a node's corresponding component is being executed, its parent node's corresponding component must have been executed as well.


\subsection{Pipeline Search Tree Algorithm} 
\label{pipeline_search_tree_algorithm}

Algorithm~\ref{alg:traverse-and-execute} outlines the traversal and execution of a pipeline search tree. In Algorithm~\ref{alg:traverse-and-execute}, $table$ denotes the compatibility LUT, $rootNode$ represents the root node of the pipeline search tree. Incompatible nodes are removed in line 5. Once the traversing reaches any leaf node, a new candidate pipeline (stored in $walkingPath$) is ready to be executed (line 15). After the execution, all the pipeline components on this path are marked as executed (lines 16-19). 
We assume that the pseudo code features \emph{passing objects by reference}, and thus the updates on nodes within $walkingPath$ will be reflected on the relevant tree node.
When a new $walkingPath$ is executed in function $executeNodeList$, \system{} can leverage $node.executed$ property to skip certain components. Let's refer back to Fig.~\ref{fig:pipeline-search-tree}. 
By leveraging the pruning heuristics, only 6 components (with orange background) corresponding to 5 pipelines, are needed to be executed.

\begin{algorithm}
\scriptsize
    \caption{Traversal and execution of the nodes on pipeline search tree with pruning heuristics.}
    \label{alg:traverse-and-execute}
    \textbf{Input}: table, rootNode\\
    \textbf{Output}: rootNode\\
    \Fn{\ExecuteTree{rootNode}}{
        \eIf{node.children $\neq \emptyset$}{
            \ForEach{child $\in$ node.children}{
                \eIf{(node.component, child.component)  $\notin$ table}{
                    node.children.remove(child)
                }
                {
                    walkingPath.push(child)\\
                    \ExecuteTree{child}\\
                    walkingPath.pop()\\
                }
            }
        }{
            executeNodeList(walking\_path)\\
            \ForEach{node $\in$ walkingPath}{
                node.executed $\gets$ True\\
                node.output $\gets$ walking\_path.getOutput[component]\\
            }
        }
    }
\end{algorithm}

\section{Evaluation}
\label{sec: evaluation}
\subsection{Evaluated Pipelines}
In this section, we evaluate the performance of \system{} in terms of storage consumption and computational time using four real-world ML pipelines, namely, readmission prediction (Readmission), Disease Progression Modeling (DPM), Sentiment Analysis (SA), and image classification (Autolearn). 
These pipelines cover different application domains such as healthcare, natural language processing, and computer vision.

\textbf{Readmission Pipeline}: The Readmission pipeline illustrated in Fig.~\ref{fig:branch-ffmerge} is built to predict the risk of hospital readmission within 30 days of discharge.
It involves three major steps: 
1) clean the dataset by filling in the missing diagnosis codes; 
2) extract readmission samples and their medical features, e.g., diagnoses, procedures, etc; 
3) train a deep learning (DL) model to predict the risk of readmission.

\textbf{DPM Pipeline}: The DPM pipeline is constructed to predict the disease progression trajectories of patients diagnosed with chronic kidney disease using the patients' one-year historical data, including diagnoses and lab test results. 
It involves four major steps where the first two steps are cleaning the dataset and extracting relevant medical features. 
In the third step, a Hidden Markov Modeling (HMM) model is designed to process the extracted medical features so that they become unbiased. 
In the last step, a DL model is built to predict the disease progression trajectory. 

\textbf{SA Pipeline}: The SA pipeline performs sentiment analysis on movie reviews. 
In this pipeline, the first three steps are designed to process the external corpora and pre-trained word embeddings. 
In the last step, a DL model is trained for the sentiment analysis task.

\textbf{Autolearn Pipeline}: The Autolearn pipeline is built for image classification of digits using Zernike moments as features. 
In the first three pre-processing steps of this pipeline, Autolearn~\cite{kaul2017autolearn} algorithm is employed to generate and select features automatically. 
In the last step, an AdaBoost classifier is built for the image classification task.

For these four pipelines, the pre-processing methods of DPM, SA, and Autolearn pipelines are costly to run, while for the Readmission pipeline, a substantial fraction of the overall run time is spent on the model training.

\subsection{Performance Metrics and Baselines}
\label{exp-baseline}
For each pipeline, we evaluate the system performance under two different scenarios: 
linear versioning and non-linear versioning.
For linear versioning performance, we perform a series of pipeline component updates and pipeline retraining operations to collect the statistics on storage and run time. 
\icde{
In every iteration, we update the pre-processing component at a probability of 0.4 and update the model component at a probability of 0.6.
At the last iteration, the pipeline is designed to have an incompatibility problem between the last two components.
} 
For the non-linear versioning performance, we first generate two branches, then update components on both branches and merge the two updated branches with the proposed version control semantics.

\textbf{Baseline for Linear Versioning}: 
\icde{
We compare \system{} against two state-of-the-art open-source systems, ModelDB~\cite{vartak2016modeldb} and MLflow~\cite{zaharia2018accelerating}. 
These two systems manage different model versions to support reproducibility.
Users are provided tracking APIs to log parameters, code, and results in experiments so that they can query the details of different models and compare them.
For these two systems, ModelDB does not offer automatic reuse of intermediate results and MLflow is able to reuse intermediate results. 
The storage mechanism of both systems archives different versions of libraries and intermediate results into separate folders.
}

\textbf{Baselines for Non-linear Versioning}: 
Two baselines are compared for the non-linear versioning scenario.
\emph{\system{} without PCPR} enumerates all the possible pipeline combinations, where \emph{PC} refers to ``Pruning using component Compatibility'', \emph{PR} refers to ``Pruning using Reusable output''. 
\emph{\system{} without PR} prunes all pipelines with incompatible components and enumerates all remaining pipeline combinations. 
\emph{\system{}} generates a pipeline tree and then prunes pipelines with incompatible components, and trained pipeline components. 

The evaluation metrics to measure the performance are cumulative execution time (CET), cumulative storage time (CST), cumulative pipeline time (CPT), and cumulative storage size (CSS). 
Execution time is the time consumption of running the computational components while storage time is the time needed for data preparation and transfer. 
Storage size refers to the total data storage used for training and storing the pipeline components and reusable outputs. 
Pipeline time refers to the sum of execution time and storage time. 
The execution time, storage time, storage size, and pipeline time are all accumulated every run during the merge operations for measuring non-linear versioning performance.

All the pipelines run on a server equipped with Intel Core-i7 6700k CPU, Nvidia GeForce GTX 980ti GPU, 16GB RAM, and 500GB SSD. 
\system{} and part of the pipeline components were implemented using Python version 3.6.8. 
Components written in C++ are complied with GCC version 5.4.0. 


\vspace{5mm}
\subsection{Performance of Linear Versioning}
\label{exp-linear-version}

\begin{figure}
    \centering
    \subfigure[Readmission]{\label{fig: lv-time-read}\includegraphics[width=.49\linewidth]{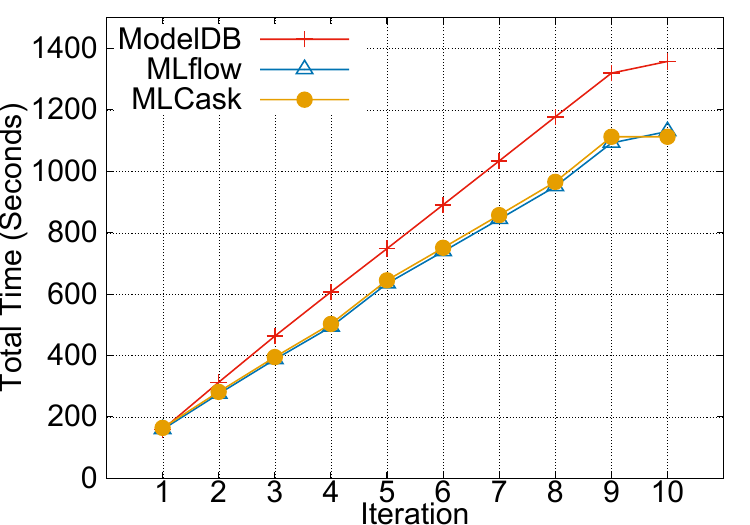}}
    \subfigure[DPM]{\label{fig: lv-time-dpm}\includegraphics[width=.49\linewidth]{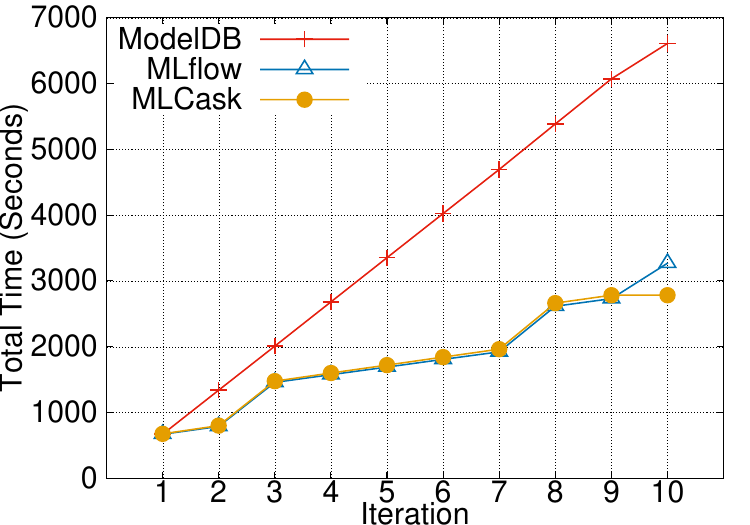}}
    \subfigure[SA]{\label{fig: lv-time-sa}\includegraphics[width=.49\linewidth]{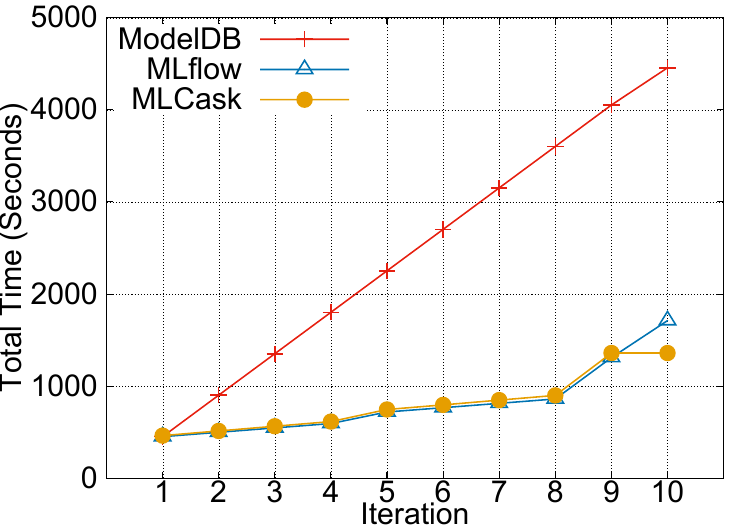}}
    \subfigure[Autolearn]{\label{fig: lv-time-autolearn}\includegraphics[width=.49\linewidth]{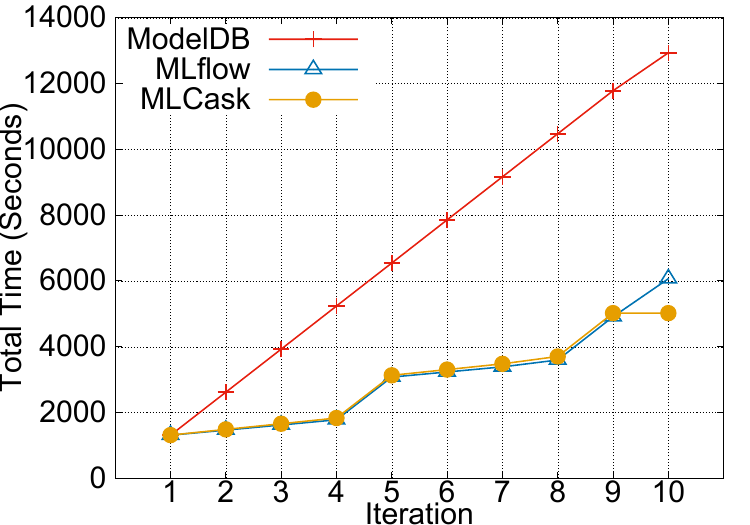}}
    \caption{\icde{Total time for linear versioning.}}
    \label{fig: lv-time}
\end{figure}


\begin{figure}
    \centering
    \includegraphics[width=0.9\linewidth]{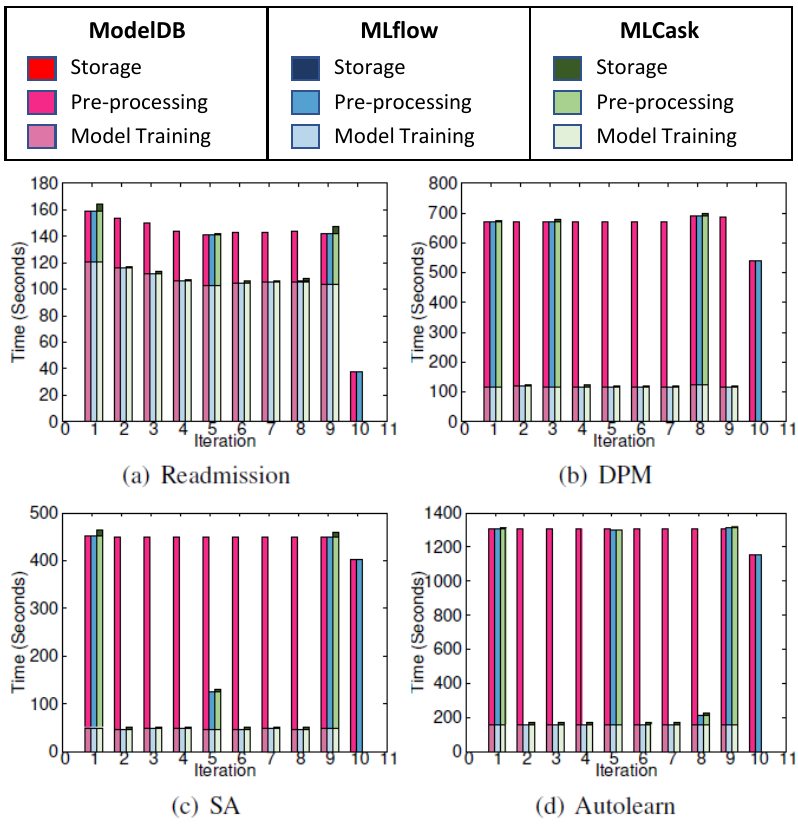}
    \caption{\icde{Pipeline time composition.}}
    \label{fig: lv-breakdown}
\end{figure}


\begin{figure}
    \centering
    \subfigure[Readmission]{\label{fig: lv-storage-read}\includegraphics[width=.49\linewidth]{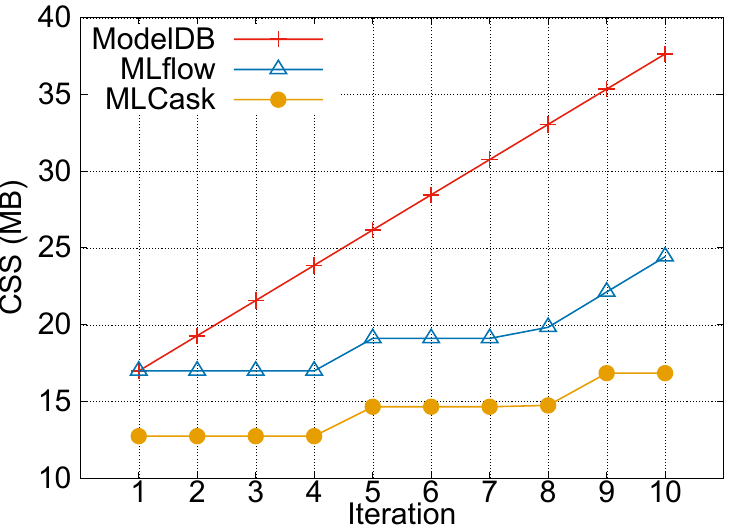}}
    \subfigure[DPM]{\label{fig: lv-storage-dpm}\includegraphics[width=.49\linewidth]{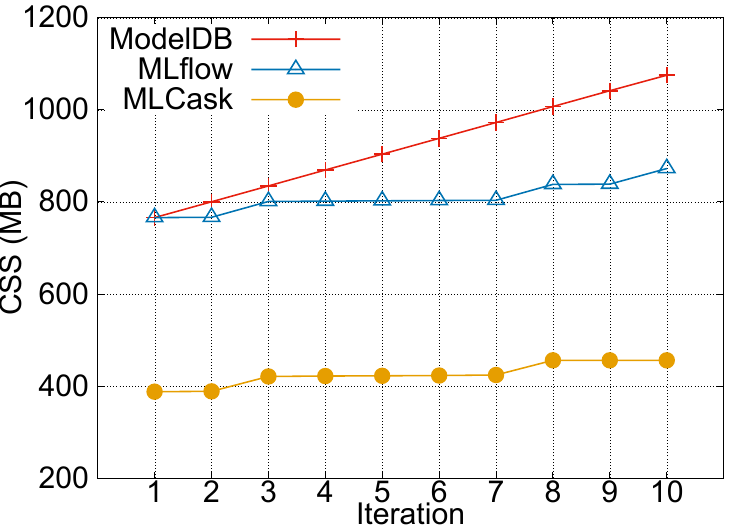}}
    \subfigure[SA]{\label{fig: lv-storage-sa}\includegraphics[width=.49\linewidth]{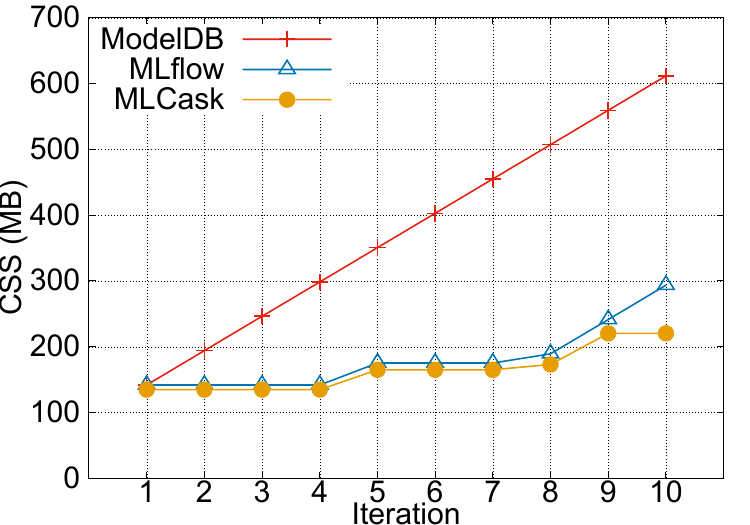}}
    \subfigure[Autolearn]{\label{fig: lv-storage-autolearn}\includegraphics[width=.49\linewidth]{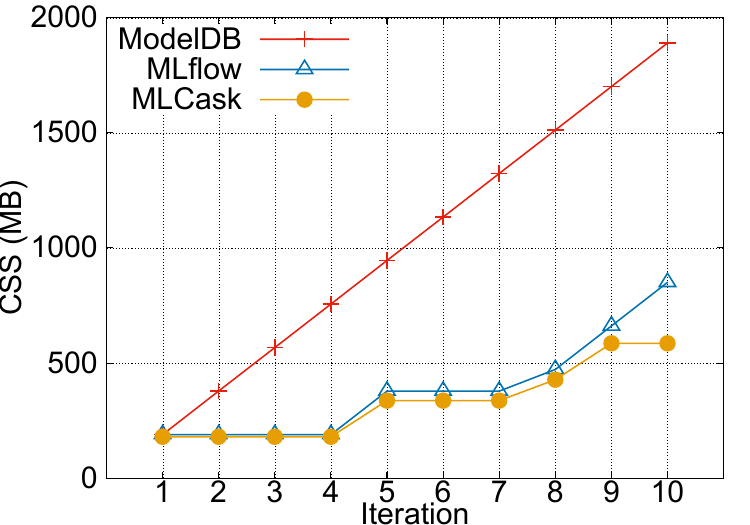}}
    \caption{\icde{Cumulative storage space for linear versioning.}}
    \label{fig: lv-storage}
\end{figure}

\icde{
Fig.~\ref{fig: lv-time} shows the total time of linear versioning on all four pipelines, and we observe that the total time of ModelDB increases linearly but at a faster rate than \system{} and MLflow in most cases. 
The linearity originates from the fact that ModelDB has to start all over in every iteration due to the lack of historical information on reusable outputs. 
\system{} and MLflow incur less pipeline time because they skip the executed pipeline components.
At the last iteration, since \system{} detects the incompatibility between the last two components before the iteration starts, it does not run the pipeline, which leads to no increase in the total time.
On the contrary, ModelDB and MLflow run the pipeline until the compatibility error occurs at the last component, which results in more pipeline time than \system{} at this iteration.

Fig.~\ref{fig: lv-breakdown} shows the pipeline time composition, and it can be observed that the time spent on model training is comparable for all systems, while the main performance difference lies in the pre-processing. 
For example, for \system{} and MLflow,  iteration 3 and iteration 8 take a longer time in the DPM pipeline. 
This is consistent with the observation from the DPM pipeline in Fig.~\ref{fig: lv-time-dpm} that the graph segment just before iterations 3 and 8 exhibits steeper slopes. 
In such cases, the updates happen on or before HMM processing, and HMM processing is time consuming, leading to a large amount of pre-processing time.
Similarly, in Fig.~\ref{fig: lv-time}, for iteration 9 of SA and iterations 5 and 9 of Autolearn, the graph segments of \system{} and MLflow exhibit steeper slopes because of the pre-processing methods, i.e., word embedding and feature generation, respectively, which can be confirmed by Fig.~\ref{fig: lv-breakdown}(c) and Fig.~\ref{fig: lv-breakdown}(d).
Specifically, for \system{} and MLflow, the pre-processing time of these iterations is significantly longer than that of other iterations.

For the storage time shown in Fig.~\ref{fig: lv-breakdown}, we note that the two baseline systems almost instantaneously
materialize the reusable outputs while \system{} takes a few seconds. 
This is because the two baseline systems store the outputs in the local directory while \system{} stores the outputs in ForkBase~\cite{wang2018forkbase} which is an immutable storage engine. 

Fig.~\ref{fig: lv-storage} shows the cumulative storage size for all the systems, and we observe that the consumption of storage by ModelDB increases linearly because every iteration is started all over and the outputs of each iteration are archived to different disk folders. 
For \system{} and MLflow, since the outputs of repeated components are stored only once and reused, these two systems consume much less storage than the ModelDB.

Further, in the first iteration, all the libraries are created and stored, and subsequently,  \system{} applies chunk level de-duplication supported by its ForkBase storage engine on different versions of libraries.
Consequently, it consumes less storage than MLflow due to its version control semantics on the libraries.
The graph segments of \system{} exhibit less steep slopes than those of MLflow for all iterations
as \system{} applies version control semantics on reusable outputs for de-duplication, 
while MLflow archives different versions of component outputs into separate folders. 

}

%
%

\subsection{Performance of Non-linear Versioning}
\label{exp-nonlinear-version}

\begin{figure}[!t]
    \centering
    \includegraphics[width=0.87\linewidth]{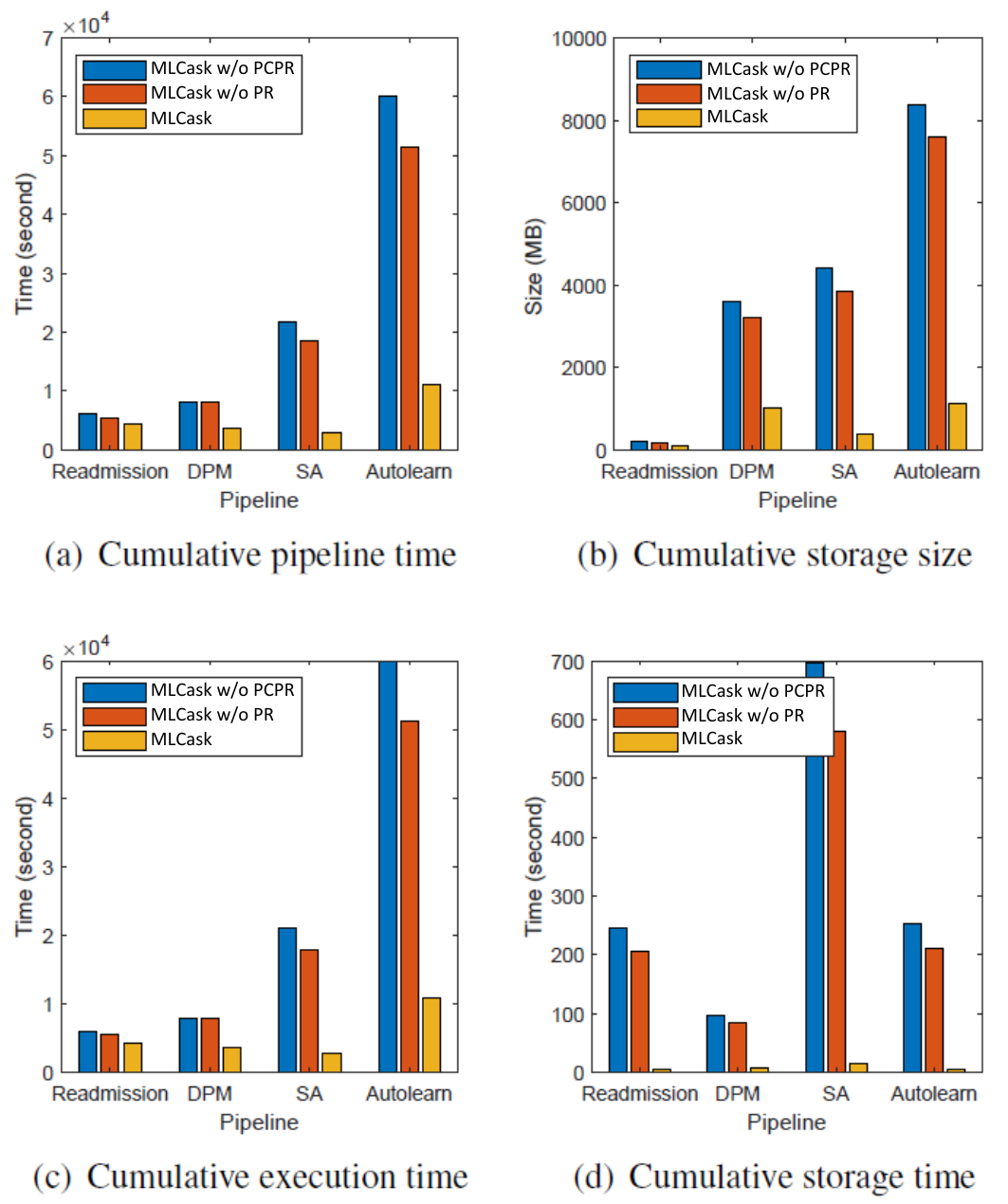}
    \caption{Non-linear versioning performance.}
    \vspace{-5mm}
    \label{fig: nl}
\end{figure}

\begin{figure}[!t]
    \centering
    \includegraphics[width=0.75\linewidth]{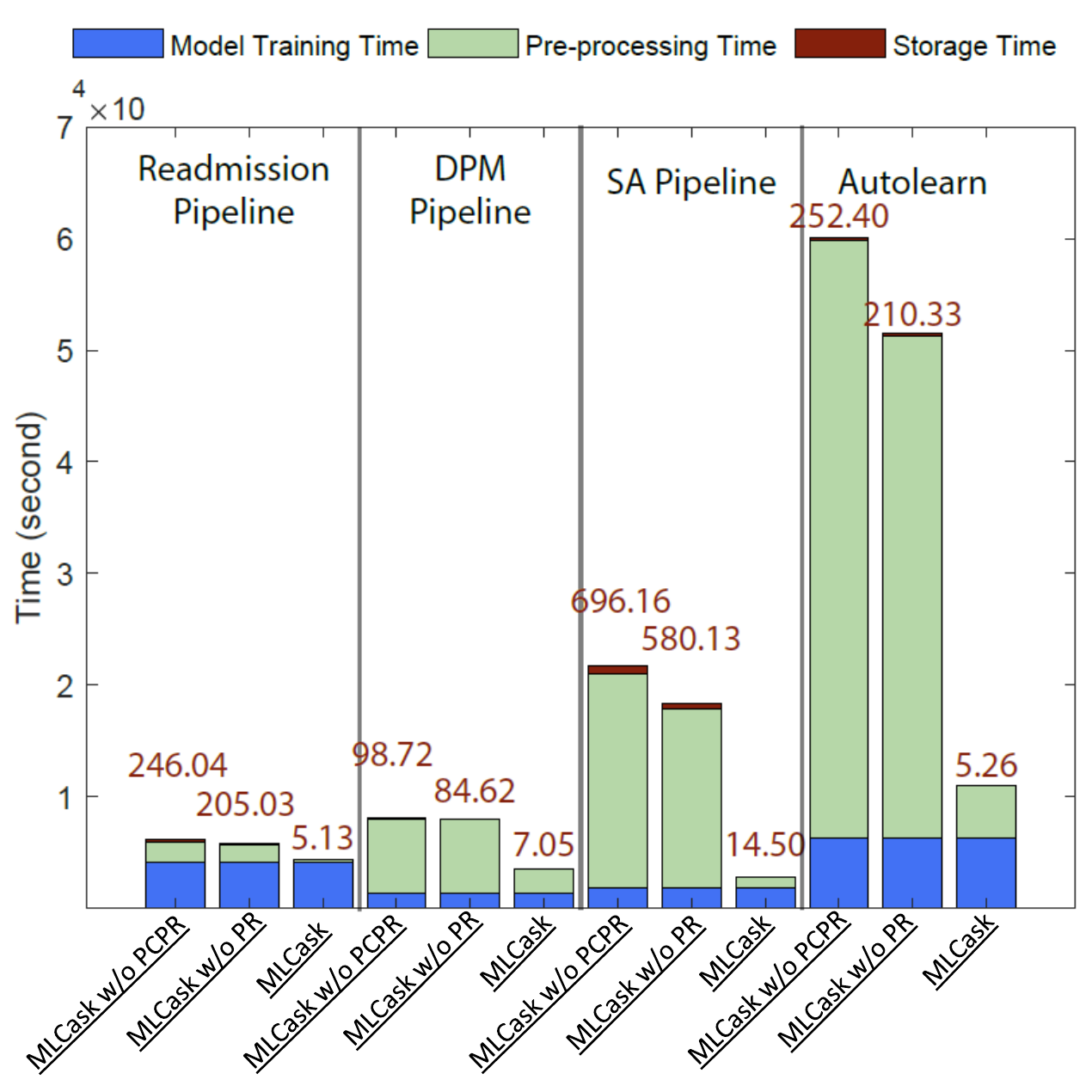}
    \caption{Pipeline time composition during merge operation.}
    \vspace{-5mm}
    \label{fig: nl-breakdown}
\end{figure}

In this section, we present the experiments on non-linear versioning, i.e., merge operation, in terms of cumulative pipeline time, cumulative storage cost, cumulative execution time, and cumulative storage time. 

Fig.~\ref{fig: nl} confirms the effectiveness in pruning the pipeline search tree using component compatibility and reusable outputs. 
The proposed system dominates the comparison in all test cases as well as all metrics, and \system{} without PR provides minor advantages over \system{} without PCPR. 

To further analyze the difference among these three systems in terms of cumulative pipeline time, we show the pipeline time composition during merge operation in Fig.~\ref{fig: nl-breakdown}. 
The difference in pipeline time among the three systems are mainly attributed to pre-processing. 
The reason is that both \emph{Pruning using component Compatibility} and \emph{Pruning using Reusable output} happen in the pre-processing components. 
For model training time, it is nearly the same across the systems. 
Storage time only constitutes a small fraction of the pipeline time.

Comparing \system{} without PR with \system{} without PCPR, \system{} without PR enumerates the possible pipelines and removes the incompatible ones explicitly before the pipeline execution, while \system{} without PCPR materializes the dataset and runs pipeline components from scratch until the compatibility error occurs.
Since the schema change happens at a lower probability, only a small subset of the pipeline candidates are removed by pruning using component compatibility. 
Consequently, the advantage of \system{} without PR over \system{} without PCPR is minor.

Comparing \system{} without PR with \system{}, the problem of \system{} without PR is that it cannot leverage the reusable outputs. 
Fig.~\ref{fig: nl} shows that this difference leads to the great advantage of \system{} over \system{} without PR.
This is because \system{} guarantees that each node on the pipeline search tree is executed only once, while for \system{} without PR, in case there are $M$ pipeline candidates, the first component in the pipeline will be executed for $M$ times. 
Therefore, the cumulative execution time and cumulative pipeline time of \system{} decrease dramatically. 

For cumulative storage size and time, Fig.~\ref{fig: nl}(b) and (d) show that \system{} outperforms the two baselines significantly because every node on the pipeline search tree is equivalent for its child nodes, and siblings of the child nodes can reuse the outputs of their parents. 
Moreover, these outputs can be stored locally as the child nodes can access the output of their parent node. 
As a result, \system{} materializes the data, typically the root node's output, and saves the final optimal pipeline (i.e., the result of merge operation) only once. 
Consequently, \system{} achieves a huge performance boost on the cumulative storage time and cumulative storage size as well.

\subsection{Prioritized Pipeline Search}
\label{exp-pps}

\begin{figure}
\centering
\includegraphics[width=0.99\linewidth]{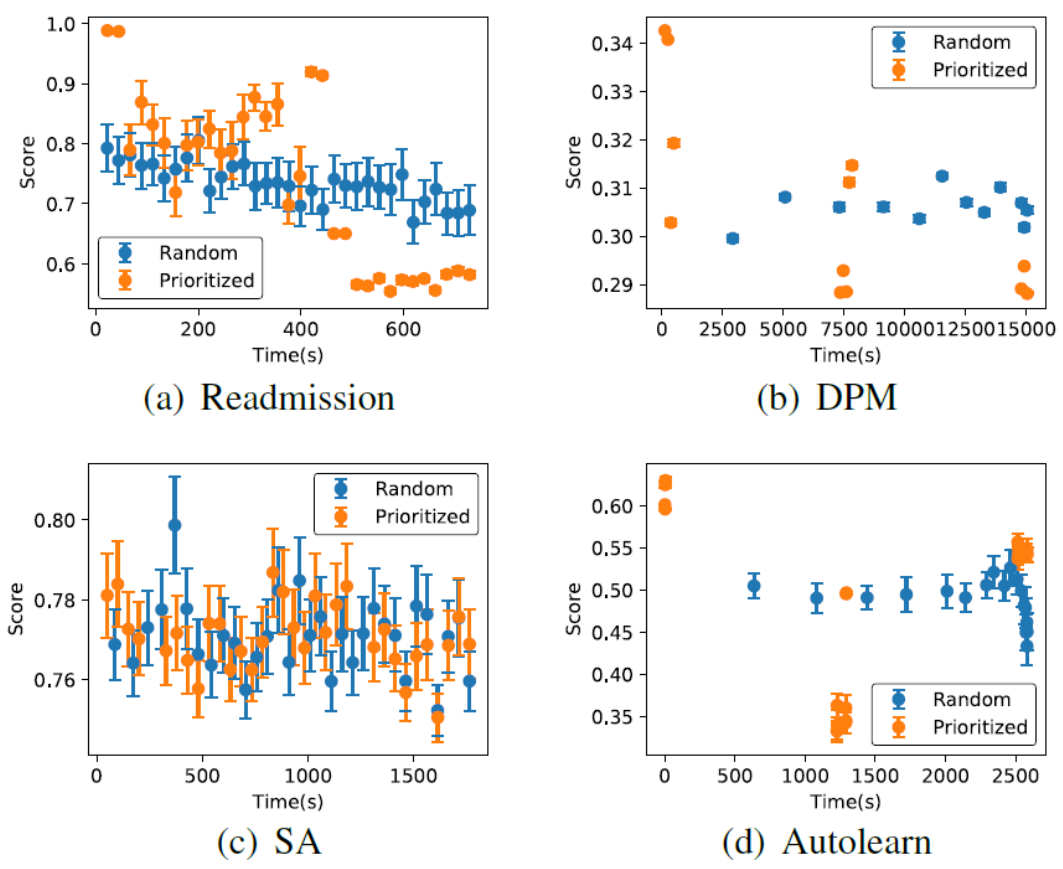}
\vspace{-4mm}
\caption{Prioritized pipeline search.}
\vspace{-6mm}
\label{fig: pps}
\end{figure}

Although pruning the pipeline search tree narrows pipeline search space, the number of pipelines that need to be evaluated may still be large.
Therefore, we prioritize the pipelines which are more promising to have desirable performance based on the pipeline history.
By doing so, the merge operation can return better results given a fixed time budget.

Every time a pipeline candidate is run, the corresponding leaf node on the pipeline search tree is associated with its score.
We associate the other nodes on the pipeline search tree with scores as well, following the rule that the score of the parent node is computed using the average of its children (except for the children that have not gotten a score yet).
The initial scores are assigned using scores of the trained pipelines on \verb|MERGE_HEAD| and \verb|HEAD|. 

Assume there are $N$ pipeline candidates (paths from the root node to the leave nodes) in the pipeline search tree.
To perform a prioritized pipeline search, we start from the root node and sequentially pick the child nodes that have the highest scores until we reach a leaf node that has not been run yet. 
This process is repeated for $N$ times so that all the $N$ pipeline candidates are searched in order.
Random search, on the contrary, searches all the $N$ pipeline candidates in random order.
For both search methods, we denote the process of searching for all the $N$ pipeline candidates in the pipeline search tree as one trial.
We perform 100 trials for both search methods and report the results in Fig.~\ref{fig: pps}.

For each application, there are $N$ points for each search method, corresponding to all the $N$ pipeline candidates.
For each point, we get the average running end time and score, as well as the variance of the scores over the 100 trials. 
It is shown that the scores obtained from prioritized search are relatively widely distributed, because the pipeline candidates searched first have higher scores while the pipeline candidates searched last have lower scores. On the contrary, the scores from random searches are nearly the same for all pipeline candidates because of the randomness. 
Meanwhile, we observe that the higher score pipeline candidates of prioritized pipeline search have a smaller average end time, which means that the high score pipeline candidates are searched first.

\shepherd{
To illustrate the effectiveness of the search strategies for finding the optimal pipeline, in Table~\ref{tbl:ratio-optimal-solution}, we show the percentage of trials in which the optimal pipeline is found after the first 20\%, 40\%, 60\%, 80\% and 100\% of searches.
When the same number of searches are conducted for each trial using each approach, the prioritized pipeline search finds more optimal pipelines than the random search because higher score pipelines are searched earlier.
Further, the optimal pipeline of all trials can be found within 80\% of
searches, and for some applications, only 40\% of searches are needed, which leads to lower computational costs.
}

\begin{table}
\centering
\scriptsize
\caption{\shepherd{Percentage of trials with the optimal pipeline found.}}
\setlength\tabcolsep{2pt}
\begin{tabular}{ccccccc} \hline
  \textbf{Application}      & \textbf{Method} & \textbf{\makecell{20\% \\ Searches}} & \textbf{\makecell{40\% \\ Searches}}  & \textbf{\makecell{60\% \\ Searches}} & \textbf{\makecell{80\% \\ Searches}}  & \textbf{\makecell{100\% \\ Searches}} \\ \hline

\multirow{2}{*}{\makecell{Read-\\mission}} & Random & 15\% & 33\% & 53\% & 78\% & 100\% \\
 & Prioritized &  \textbf{26\%} &  \textbf{49\%} &  \textbf{92\%} & \textbf{100\%} & 100\% \\
 \hline
 
\multirow{2}{*}{DPM} & Random & 23\% & 42\% & 59\% & 83\% & 100\% \\
 & Prioritized &  \textbf{44\%} &  \textbf{100\%} &  \textbf{100\%} & \textbf{100\%} & 100\% \\ \hline
 
\multirow{2}{*}{SA} & Random & 13\% & 31\% & 51\% & 72\% & 100\% \\
 & Prioritized &  \textbf{30\%} &  \textbf{84\%} &  \textbf{100\%} & \textbf{100\%} & 100\% \\ \hline
 
\multirow{2}{*}{\makecell{Auto-\\learn}} & Random & 31\% & 54\% & 78\% & 93\% & 100\% \\
 & Prioritized &  \textbf{54\%} &  \textbf{100\%} &  \textbf{100\%} & \textbf{100\%} & 100\% \\ \hline

\end{tabular}
\label{tbl:ratio-optimal-solution}
\end{table}

\shepherd{
In summary, \system{} supports two pipeline search approaches: (i) optimal approach with pruning (Section \ref{pipeline_search_tree_algorithm}), and (ii) prioritized pipeline search. 
Note that the prioritized pipeline search
only changes the order of searches without changing the search space.
Consequently, if the time budget is sufficiently large for searching all the solutions, the prioritized pipeline search will guarantee to obtain the optimal pipeline as the optimal approach with pruning does.
Nevertheless, if the time budget is limited, the prioritized pipeline search only searches the most promising pipelines according to the history.
Empirically, it is shown that the prioritized pipeline search finds better pipelines than random search under limited time budget, and the optimal pipeline is more likely to be found in the early searches. 
}

\subsection{Distributed Training on Large ML Model}
\label{sec: dist-train}
Analytics models such as DL models in the pipeline require long training time.
In this case, since \system{} supports any executable as a pipeline component, distributed training can be applied as long as the executable contains the library for distributed training. 

\begin{figure}[!t]
    \centering
    \subfigure[Training loss vs time]{\label{fig:dist_train_loss}\includegraphics[width=.49\linewidth]{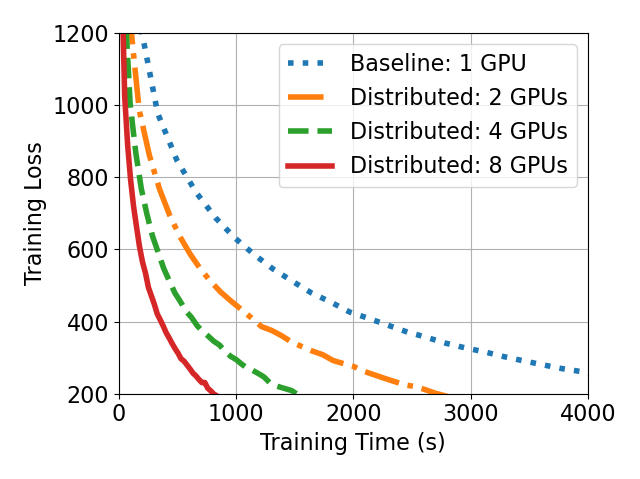}}
    \subfigure[Pipeline time speedup]{\label{fig:dist_time_speedup}\includegraphics[width=.49\linewidth]{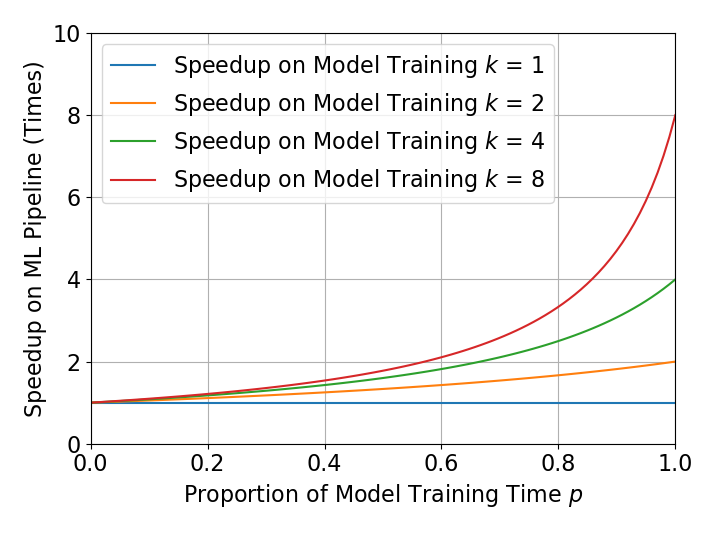}}
    \caption{Distributed training.}
    \label{fig: dist-train}
    \vspace{-5mm}
\end{figure}

In this section, we analyze how much speedup we could achieve if we apply up to 8 GPUs for synchronous distributed training in the same computing node. 
We take ResNet18~\cite{he2016deep} model as an example. 
The speedup on the model due to distributed training is shown in Fig.~\ref{fig:dist_train_loss}. 
We observe that the training loss decreases faster over training time for more GPUs. This is because more GPUs lead to an increase in sample processing throughput.
Consequently, with distributed training for the large ML models in the pipeline, it is possible that the pipeline time can be greatly reduced.

Since pipeline time consists of model training time, pre-processing time, and storage time, analyzing the pipeline time speedup brought about by distributed model training needs to take other components in the pipeline into consideration. 
We thus formalize the pipeline time speedup due to the distributed model training as: $\mathrm{\textbf{Speedup}\,} = {1}/[(1-p)+{p}/{k}] $,
where $p$ is the portion of model training time out of the total pipeline time, and $k$ is the speedup of the model training due to distributed training. 
The pipeline time speedup for different combinations of $k$ and $p$ is shown in Fig.~\ref{fig:dist_time_speedup}.
We note that both increased $k$ and increased $p$ lead to increased pipeline time speedup. 
As long as $k$ is larger than 1, the pipeline time speedup is larger than 1.
Specifically, when the portion of model training time is more than 0.9 and the speedup of the model training equals 8, the pipeline time is less than one-fourth of the original pipeline time, which saves a lot of time.


\section{Discussion on System Deployment}
\label{sec: discussion}


\icde{
In this section, we share our experience on the deployment of \system{} at National University Hospital\footnote{https://www.nuh.com.sg -- a major public and tertiary hospital in Singapore, which is part of NUHS}(NUH).
We have been working with NUHS\footnote{https://en.wikipedia.org/wiki/National\_University\_Health\_System} since 2010 on data cleaning, data integration, modeling and predictive analytics for various diseases~\cite{ling2014gemini,zheng2017resolving,zheng2020tracer},
as a collaboration to develop solutions for existing and emerging healthcare needs. 
Due to the sensitivity of the data and the critical nature of healthcare applications,  hospitals must manage the database and model development for accountability and verifiability purposes. \system{} has been designed towards fulfilling such requirements.

In deployment, the production pipeline has to be separated from the development pipeline. 
The production pipeline is a stable version that should not be modified when it is in service, unless minor bug fixes are required.
For development purposes, we form a branch with a replica of the pipeline as a development pipeline. 
For upgrading of the production pipeline, we can merge the development pipeline into the production pipeline. 
To facilitate such development and upgrading, \system{} provides branching functionality for the pipelines.

In a large hospital such as NUH, different data scientist teams and clinicians may develop models of the same pipeline concurrently. 
The scenario is similar to what has been depicted in Fig.~\ref{fig:branch-merge} and explained in Section~\ref{sec: supporting non-linear version history}, where different users are updating different components of the same pipeline at the same time. 
This could lead to a number of updated pipelines that are difficult to be merged together. 
As explained in Section~\ref{sec: supporting non-linear version history}, using a na\"ive strategy to select the latest components could lead to incompatibility and sub-optimal pipeline issues. 
To this end, \system{} supports pipeline merging optimization to derive a more effective pipeline.

In summary, \system{} has been designed to address three issues encountered in a hospital deployment: (i) frequent retraining, (ii) needs for branching, and (iii) merging of updated pipelines.
Apart from NUH, \system{} is being adapted for another major public hospital in Singapore.
}
\section{Related Work}
\label{sec: related work}
\noindent \textbf{Versioning for Datasets and Source Code.} 
State-of-the-art systems for managing datasets versioning such as Forkbase \cite{wang2018forkbase}, OrpheusDB \cite{huang2017rpheus}, and Decibel \cite{maddox2016decibel} support \git{}-like semantics on datasets to enable collaborative analysis as well as efficient query processing. 
In terms of versioning code of pre-processing methods and models, the file-based \git{} is widely used. They store source code in repositories and manage versions of the code based on the text information. 
However, these methods are not suitable for managing the versioning of the data analytics pipeline.
Compared with dataset versioning, pipeline versioning requires not only dataset versioning but also the versioning of the source code. 
Furthermore, in contrast to \git{}, pipeline versioning needs to take care of the evolution of the whole pipeline, which comprises the source code, the datasets, and the relationship between pipeline components.

\noindent \textbf{Build Automation Tools.} 
In terms of maintaining the relationships between pipeline components, build automation tools such as 
Maven\footnote{Apache Maven: http://maven.apache.org}, Gradle\footnote{Gradle: http://gradle.org} and Ant\footnote{Apache Ant: http://ant.apache.org} manage the dependency between different software packages to facilitate the project development. 
In comparison, \system{} has a quite different objective: pipeline versioning organizes various subsystems to form an end-to-end data analytics pipeline instead of compiling a project.
Further, pipeline versioning requires explicit data-flow management to enable the saving or reusing of the intermediate outputs for exploration, which is not an objective of the build automation tools. 

\noindent \textbf{Data Version Control (DVC).}
DVC\footnote{DVC: https://dvc.org} is a system built upon \git{}, which supports non-linear version history of pipelines, and also records the performance of the pipelines.
Unfortunately, it inherits the merge mechanism from \git{}, which treats merge operation as combining the latest features.

\noindent \textbf{Machine Learning Pipeline Management.} 
In ML pipeline management, MLlib~\cite{JMLR:v17:15-237} simplifies the development of ML pipelines by introducing the concepts of DataFrame, Transformer, and Estimator. 
SHERLOCK~\cite{vartak2015supporting} enables users to store, index, track, and explore different pipelines to support ease of use, while Velox~\cite{crankshaw2014missing} focuses on online management, maintenance, and serving of the ML pipelines. 
Nevertheless, version control semantics of the pipelines are not supported by the aforementioned methods.

The pipeline management system that is most similar to \system{} is proposed in \cite{van2017versioning}. In this work, versioning is proposed to maintain multiple versions of an end-to-end ML pipeline.
It archives different versions of data into distinctive disk folders, which may lead to difficulty in tracing the version history and incur a huge storage cost. 
This work addresses the asynchronous pipeline update problem. 
However, how to set the version number remains undefined. 

Another line of research works focuses on using intermediate results for optimizing the execution of ML pipelines or for diagnosis.
\icde{
ModelDB~\cite{vartak2016modeldb} and MLflow~\cite{zaharia2018accelerating} provide a tracking API for users to store the intermediate results to a specific directory.
}
Helix~\cite{xin2018helix} reuses intermediate results as appropriate via the
Max-Flow algorithm. 
Derakhshan et al.~\cite{derakhshan2020optimizing} materialize the intermediate results that have a high likelihood of future reuse and select the optimal subset of them for reuse.
For debugging or diagnosing the ML pipelines,
MISTIQUE~\cite{vartak2018mistique} efficiently captures, stores, and queries intermediate results for diagnosis using techniques such as quantization, summarization, and data de-duplication. 
Zhang et al.~\cite{zhang2017diagnosing} diagnose the ML pipeline by using fine-grained lineage, e.g., elements in a matrix or attributes in a record. 
The above mentioned works emphasize the use of intermediate results as opposed to addressing the non-linear version history problem.


\section{Conclusions}
\label{sec: conclusions}
In this paper, we propose \system{} to address the key challenges of constructing an end-to-end \git{}-like ML system for collaborative analytics, in the context of developing or maintaining data analytics applications. 
Firstly, non-linear pipeline version control is introduced to isolate pipelines for different user roles and various purposes.
Secondly, the challenge of the asynchronous pipeline update is supported with lineage tracking based on semantic versioning and the ML oriented merge operation.
Thirdly, two pruning methods are proposed to reduce the metric-driven merge operation cost for the pipeline search.
For a resource efficient solution under a limited time budget, we present the prioritized pipeline search which provides the trade-off between time complexity and solution quality.
Extensive experimental results confirm the superiority of \system{} in terms of storage cost and computation efficiency.
\system{} has been fully implemented and deployed at a major public hospital.
\section*{Acknowledgment}
This research is supported by the National Research Foundation Singapore under its AI Singapore Programme (Award Number: AISG-GC-2019-002). 
Meihui Zhang's work is supported by National Natural Science Foundation of China (62050099).
\bibliographystyle{abbrv}
\bibliography{ref-clean}

\end{document}

%% file: terminology.tex
\newcommand*{\system}{MLCask}
\newcommand*{\git}{Git}

\newcommand*{\hospital}{NUHS}